\documentclass[aps,prb,twocolumn,showpacs,floatfix,superscriptaddress]{revtex4-1}
\usepackage{graphicx,color}
\usepackage{amsfonts}
\usepackage[figuresright]{rotating}
\usepackage{amssymb}
\usepackage{amsmath}
\usepackage{psfrag}
\usepackage{subfigure}
\usepackage{multirow}
\usepackage{tabularx}
\usepackage{textcomp}
\usepackage{units}
\usepackage{hyperref}
\hypersetup{
 pdfnewwindow=true, colorlinks=true,
 linkcolor=blue, anchorcolor=blue,
 citecolor=blue, filecolor=blue,
 menucolor=blue, urlcolor=blue}

\usepackage{graphicx}
\usepackage{dcolumn}
\usepackage{bm}
\usepackage{color}

\makeatletter

\begin{document}


\title{Evolution of elastic moduli through a two-dimensional structural transformation}

\author{Alejandro\ \surname{Pacheco-Sanjuan}}
\email{alejandro.pachecos@usm.cl}
\affiliation{Departamento de Ingenier{\'\i}a Mec\'anica, Universidad T\'ecnica Federico Santa Mar{\'\i}a, Valpara{\'\i}so, Chile}
\author{Tyler B.\ \surname{Bishop}}
\affiliation{Department of Physics, University of Arkansas, Fayetteville, AR 72701, USA}
\author{Erin E.\ \surname{Farmer}}
\affiliation{Department of Physics, University of Arkansas, Fayetteville, AR 72701, USA}
\author{Pradeep\ \surname{Kumar}}
\affiliation{Department of Physics, University of Arkansas, Fayetteville, AR 72701, USA}
\author{Salvador\ \surname{Barraza-Lopez}}
\email{sbarraza@uark.edu}
\affiliation{Department of Physics, University of Arkansas, Fayetteville, AR 72701, USA}
\affiliation{Institute for Nanoscience and Engineering, University of Arkansas, Fayetteville, AR 72701, USA}
\begin{abstract}
We use a classical analytical and separable elastic energy landscape describing SnO monolayers to estimate the softening of elastic moduli {through a mechanical instability occurring at finite temperature in this material}. Although not strictly applicable to this material due to its low energy barrier $J$ that leads to a quantum paraelastic phase, the present exercise is relevant as it establishes a conceptual procedure to estimate such moduli straight from a two-dimensional elastic energy landscape. As additional support for the existence of a quantum paraelastic phase, we carry a qualitative WKB analysis to estimate escape times from an individual well on the landscape; escape times increase exponentially with the height of the barrier $J$. We also provide arguments against an additional transformation onto a planar lattice due to its high energy cost. These results continue to establish a case for the usefulness of soft matter concepts in two-dimensional materials, and of the potential lurking of quantum effects into soft matter.
\end{abstract}

\date{\today}

\maketitle

\section{Introduction}


The earliest indication of structural transformations of two-dimensional materials dates back to 1996, when a structural phase transition driven by an electric field was demonstrated at the surface layer of TaSe$_2$ nanocrystals.\cite{tmdm1,tmd0} Similar transitions have been achieved in MoTe$_2$ monolayers recently.\cite{tmd1,tmdelectrostatic}

Group-IV monochalcogenide monolayers ({\em e.g.}, SnSe and SnTe) and SnO monolayers were introduced as potential two-dimensional materials that undergo structural transformations driven either by temperature,\cite{Mehboudi2016} strain and/or charge doping.\cite{Seixas2016} The thermally-driven structural transformation of monochalcogenide monolayers has been experimentally verified.\cite{Kai} Additional experiments have enlarged the number of 2D materials displaying ferroic behavior,\cite{cips,in2se3_1,in2se3_2,in2se3_3,in2se3_4,in2se3_4,wte2,kai2,snseml,sutter} and theory continues to increase the potential functionalities of these materials ({\em e.g.}, Refs.~\onlinecite{other1,other2,other3,other4,review,b1,b2,b3,b4,new,rev}, to mention a few).

\begin{figure}[tb]
\begin{center}
\includegraphics[width=0.48\textwidth]{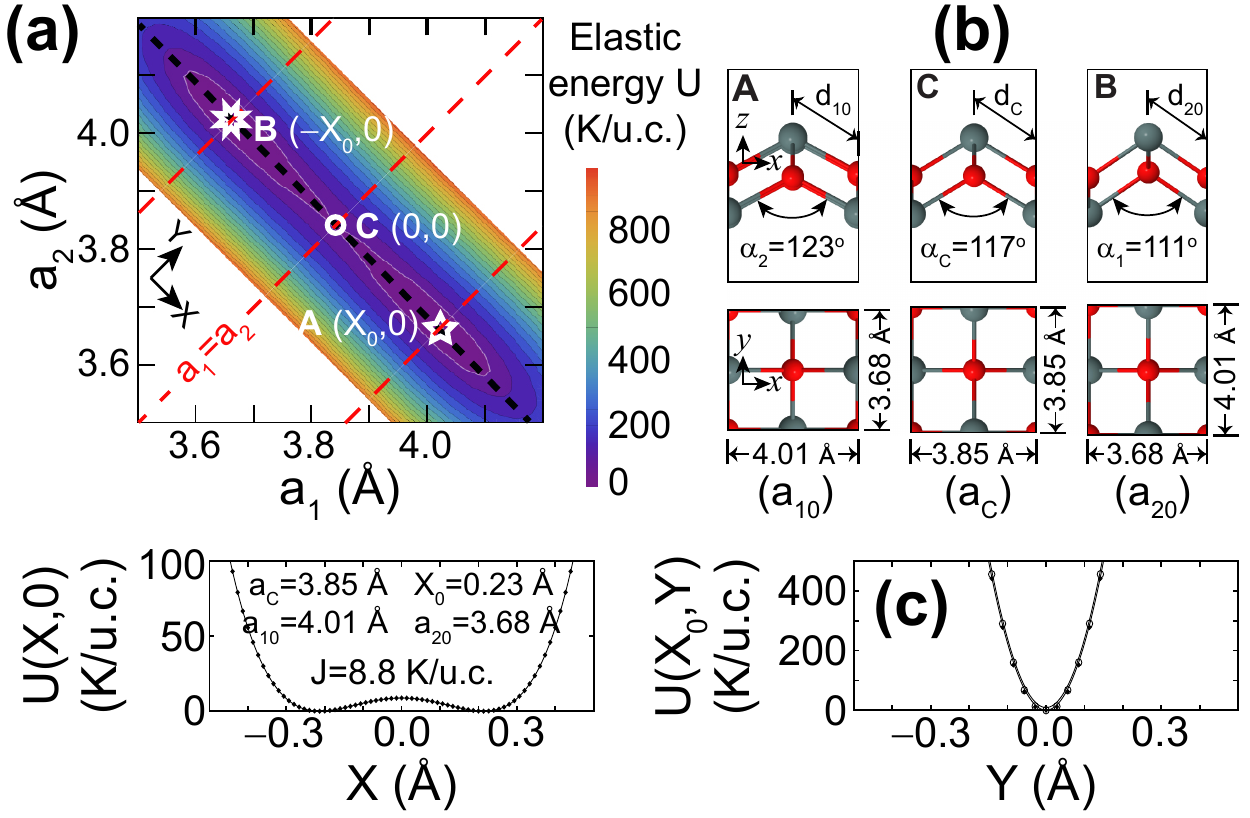}
\caption{(a) Elastic energy landscape for the SnO monolayer with zero doping. (b) Unit cells at the energy minimum (structures $A$ and $B$) and for the square structure at point $C$. Structural order parameters are shown. (c) Energy cuts through the black and red dashed lines shown in subplot (a). (This figure is a reproduction of Fig.~1 in Ref.~\onlinecite{arxiv} that is presented here for self-completeness.)}\label{fig:f0}
\end{center}
\end{figure}

We revisited the structure {\em versus} temperature properties SnO monolayers using {\em ab initio} calculations of unit cells at zero temperature, and {\em ab initio} molecular dynamics calculations of charge neutral SnO supercells at finite temperature {recently.\cite{arxiv} To make the present work self-contained, the structure of the SnO monolayer and its energy landscape originally presented in Ref.~\onlinecite{arxiv} are shown in Fig.~\ref{fig:f0} and briefly discussed next.

Figure \ref{fig:f0}(a) shows the total energy of a SnO monolayer as a function of its two orthogonal lattice vectors $\mathbf{a}_1=(a_1,0,0)$ and $\mathbf{a}_2=(0,a_2,0)$. Such energy is shown relative to the two local minima located at points $A$ and $B$, which are related by an exchange of lattice vectors are are hence degenerate, and indicated in units of Kelvin per unit cell (K/u.c.). These calculations were performed with the {\em VASP} code\cite{vasp} within the PBE approximation for exchange and correlation\cite{PBE} and employed PAW pseudopotentials\cite{paw,vasppseudos}. A $15\times 15 \times 1$ Monkhorst-Pack \cite{mp} $k-$point mesh centered about the $\Gamma-$point, and a 500 eV energy cutoff for the plane wave expansion were employed. Once a pair of values for $a_1$ and $a_2$ were chosen, a structural optimization of the basis atoms was performed with fixed lattice vectors until atomic forces became smaller than $10^{-3}$ eV/\AA; $\mathbf{a}_3=(0,0,10$ \AA$)$ in all calculations.

As indicated in Ref.~\onlinecite{arxiv}, the degenerate ground-state structure $A$ ($B$) in Fig.~\ref{fig:f0}(b) has lattice parameters $a_{10}=4.01$ \AA{} and $a_{20}=3.68$ \AA{} ($a_{10}=3.68$ \AA{} and $a_{20}=4.01$ \AA{}), distances among oxygen and tin atoms $d_{10}$ and $d_{20}$ are 2.28 and 2.23 \AA, respectively, and the two angle formed among two tin atoms and an oxygen atom situated along the $x-$direction ($y-$direction) is 123$^{\circ}$ (111$^{\circ}$).

Point $C$ in Fig.~\ref{fig:f0}(a) lies midway points $A$ and $B$ at $a_1=a_2=a_c=3.85$ \AA. The distance among oxygen and tin atoms is $d_C=2.25$ \AA{}, and the angle among an oxygen and two consecutive tin atoms is 117$^{\circ}$ in this structure. $J$ is the energy barrier, defined as the difference among the energy at degenerate point $A$ (or $B$) and that at point $C$. $J$ is the lowest energy needed to swap the crystal in between $A$ and $B$ configurations,  and its magnitude is a mere 8.8 K/u.c. In the structural transformation being considered here, coordination remains fourfold, and macroscopic monodomains with configurations $A$ or $B$ turn onto structure $C$.}

{Figure \ref{fig:f0}(c) displays two orthogonal cuts of the landscape, Fig.~\ref{fig:f0}(a) along the diagonal lines $X=(a_1-a_2)/\sqrt{2}$ and $Y=(a_1+a_2-2a_c)/\sqrt{2}$ shown by black and red dashed lines, respectively. As indicated before, the energy dependency along the $Y$ direction is parabolic, and has a small dependency on $X$. In terms of variables $X$ and $Y$, structure $C$ lies at $(0,0)$ while structures $A$ or $B$ are at $(\pm X_0,0)$, respectively ($X_0=0.23$ \AA).\cite{arxiv}}

{In our previous work,} we concluded that ferroic behavior should not be expected when the elastic energy barrier $J$ ({\em i.e.}, the energy difference between the degenerate structural ground states and the unit cell with enhanced symmetry that is mid-way among {\em all} degenerate ground states) is of the order of a few tens of Kelvin per unit cell. For in that situation, Bose-Einstein statistics lead to quantum fluctuations large enough for atoms to overcome the energy barrier and co-populate the two minima in the energy landscape, in a phenomena called quantum paraelasticity.

In the present manuscript, we continue our study of neutral SnO monolayers, and provide additional analysis and techniques that could be useful for further studies of the elastic properties of 2D materials at the onset of structural transformations. As the main result, and relying on the simplicity of the energy landscape, we consider in Sec.~\ref{sec:1} the classical evolution of elastic moduli across a 2D transformation whose elastic energy landscape was defined analytically. Even though paraelasticity may render this particular analysis irrelevant for SnO, such type of studies are desirable within the context of soft matter and statistical physics that make use of 2D models,\cite{Gerardo1,Mao2015,Lub} and have value from a model perspective.

Two additional topics are discussed briefly afterwards. In Sec.~\ref{sec:3}, we employ a textbook example to facilitate a second argument for a quantum paraelastic phase on SnO monolayers: considering the two wells on the analytic elastic landscape, we use the WKB approximation to estimate the escape time of a particle --with mass $m$ equal to that of the four atoms in the unit cell-- off an individual well. We document an exponential increase on the escape time as the barrier height is increased on the analytical model.

Once the litharge structure is achieved (in which the two orthogonal lattice vectors have equal lattice parameters $a_1=a_2$), there is still another possible two-dimensional transformation in which the unit cell turns planar. A second energy barrier, $J'>J$ is presented and its consequences discussed in Sec.~\ref{sec:4}. Conclusions are provided afterwards.

{While a revision of the present paper was written, we learned of prior work on this subject carried out by Zhong and Vanderbilt on bulk SrTiO$_3$ and BaTiO$_3$ \cite{zhong} and by Lebedev \cite{lebedev1} on few-layer SnS, where the effect of quantum fluctuations has been studied. In particular, Zhong and Vanderbilt implemented a path-integral quantum Monte Carlo framework on an analytical elastic energy landscape to estimate the effects of quantum fluctuations quantitatively. Our approach is rather qualitative in comparison, but it begins to open up the existence of similar effects in two-dimensional ferroelectrics and adds a number of qualitative arguments for quantum paraelasticity.}

\section{Elastic properties from the analytic energy density}\label{sec:1}
Working on a model of a 2D elastic media, Mao and coworkers state that structural transitions are signalled by the softening of phonon modes at discrete points in the Brillouin zone --something we recently observed in SnO monolayers\cite{arxiv}-- and therefore by the softening of certain elastic moduli.\cite{Mao2015} The SnO monolayer has a coordination number $z=4$, placing it at the edge of mechanical instability, given that $z=2d$ and $d=2$ for this two-dimensional lattice.\cite{Gerardo1,Mao2015}

Elastic moduli $C_{ijkl}$ are usually defined in terms of Gibbs free energy as follows:\cite{Batra,Landau}
\begin{equation}\label{eq:constants}
C_{ijkl}=\frac{1}{\Omega_0}\frac{\partial^2(\Psi+P\Omega)}{\partial \epsilon_{ij}\partial \epsilon_{kl}},
\end{equation}
where $\Omega_0$ ($\Omega$) is the volume at zero (finite) temperature, $\Psi$ is Helmholtz free energy, and $P$ stands for pressure. We estimate an energy contribution of the order of 10 mK/u.c. from the $P\Omega$ term at ambient pressure, and thus follow the standard practice of disregarding this term in what follows.

Then, we approximated the landscape so that it is separable on $X$ and $Y$; details are given in Ref.~\onlinecite{arxiv}. This separable energy landscape permits estimating thermodynamical averages of any function of the landscape's coordinates analytically. Such approach was employed in previous work to estimate the evolution of lattice parameters,\cite{arxiv} but more complex functions can be evaluated as well, and we analyze the thermally-induced softening of elastic moduli next,\cite{Landau,Batra} something we have not seen done within the context of 2D materials thus far.

The elastic energy landscape of the charge-neutral SnO monolayer looks as follows:\cite{arxiv}
\begin{equation}\label{eq:U}
U(X,Y)=\frac{b^2}{4a}+aX^4-bX^2+cY^2,
\end{equation}
where {$a=4252$ K\AA$^{-4}/u.c.$, $b=387$ K\AA$^{-2}/u.c.$, and $c=22703$ K\AA$^{-2}/u.c.$ are obtained as a fit against the landscape in Fig.~\ref{fig:f0}(a). Variables $a$ and $b$ set a double well potential along the $X$ direction, and $c$ provides a harmonic dependence on $Y$. Note that $X_0=\sqrt{b/(2a)}$ and $J=U(0,0)=b^2/(4a)$.\cite{arxiv}}

If a sufficiently large monodomain exists\cite{review} --characterized by a sizeable number of unit cells whose lattice vectors are ($a_{10}$,0,0) and (0,$a_{20}$,0)-- then the contribution of domain walls to the elastic energy can be initially omitted, and unitary cartesian displacements $\epsilon_{11}$ and $\epsilon_{22}$ along the $a_1-$ and $a_2-$directions can be expressed around the monodomain minima having unequal lattice constants (coordinates) $a_{10}$ and $a_{20}$ at point $A$ in Fig.~\ref{fig:f0}:
\begin{equation}
\epsilon_{11}=\frac{a_1-a_{10}}{a_{10}}, \text{ }\epsilon_{22}=\frac{a_2-a_{20}}{a_{20}};
\end{equation}
{elastic properties are usually expressed against such unitary displacements, prompting such reparametrization of the energy landscape.}
The definition of the displacements with respect to the minima is at variance of Ref.~\onlinecite{Seixas2016}, where they are expressed with respect to (unstable) point $C$.

We now express lattice constants in terms of unitary displacements $a_1=a_{10}(1+\epsilon_{11})$, and $a_2=a_{20}(1+\epsilon_{22})$, so that $X(a_1,a_2)$ and $Y(a_1,a_2)$ in Ref.~\onlinecite{arxiv} become:
\begin{eqnarray}\label{eq:eq3}
X(\epsilon_{11},\epsilon_{22})&=\frac{a_{10}(1+\epsilon_{11})-a_{20}(1+\epsilon_{22})}{\sqrt{2}},\text{ and}\nonumber\\
Y(\epsilon_{11},\epsilon_{22})&=\frac{a_{10}\epsilon_{11}+a_{20}\epsilon_{22}}{\sqrt{2}},
\end{eqnarray}
and the elastic energy per unit cell turns into:
\begin{eqnarray}
U(\epsilon_{11},\epsilon_{22})=\frac{b^2}{4a}+\frac{a}{4}\left[a_{10}(1+\epsilon_{11})-a_{20}(1+\epsilon_{22})\right]^4\nonumber\\
-\frac{b}{2}\left[a_{10}(1+\epsilon_{11})-a_{20}(1+\epsilon_{22})\right]^2+\frac{c}{2}\left[a_{10}\epsilon_{11}+a_{20}\epsilon_{22}\right]^2.\nonumber
\end{eqnarray}

\begin{figure}[tb]
\begin{center}
\includegraphics[width=0.48\textwidth]{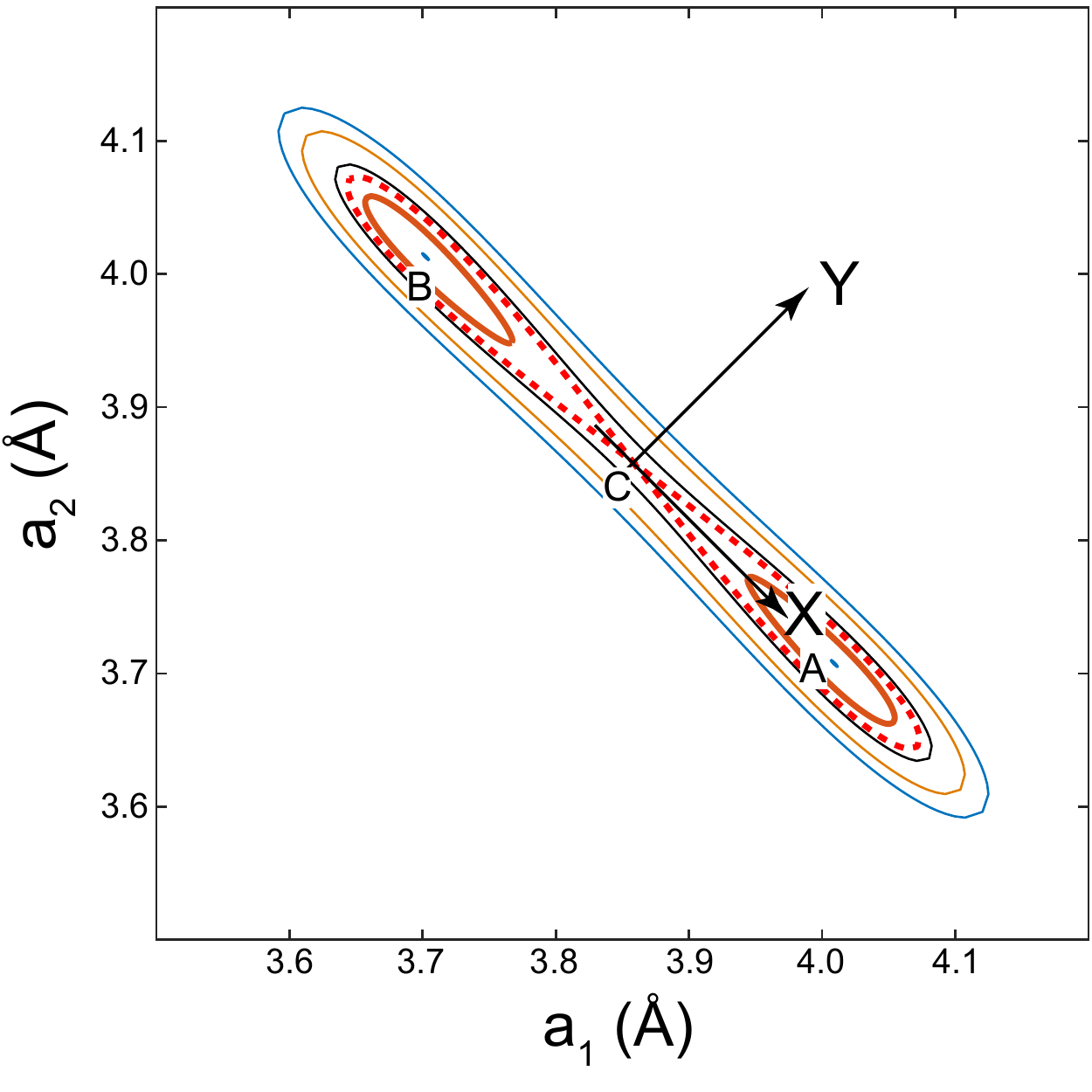}
\caption{Isoenergy contours on the two-dimensional landscape (Fig.~\ref{fig:f0}(a)) as parameterized by Eq.~\ref{eq:U}. The red dashed curve corresponds to $k_BT=J$ (per unit cell). Isoenergy paths $k_BT<J$ are disconnected, implying that a particle around minima $A$ (with coordinates ($a_{10}$, $a_{20}$)) does not have enough kinetic energy to jump onto the valley where minima $B$ ($a_{20}$, $a_{10}$)) resides. Coordinates $X$ and $Y$ are shown too; their origin is at point ($a_C$,$a_C$), with $a_C=(a_{10}+a_{20})/2$.}\label{fig:f1}
\end{center}
\end{figure}

A number of derivatives are needed to express these moduli:
\begin{eqnarray}\label{eq:eq4}
\frac{1}{a_{10}}\frac{\partial U}{\partial \epsilon_{11}}=a\left[a_{10}(1+\epsilon_{11})-a_{20}(1+\epsilon_{22})\right]^3\text{ }\\
-b\left[a_{10}(1+\epsilon_{11})-a_{20}(1+\epsilon_{22})\right]+c\left[a_{10}\epsilon_{11}+a_{20}\epsilon_{22}\right],\nonumber\\
\frac{1}{a_{20}}\frac{\partial U}{\partial \epsilon_{22}}=-a\left[a_{10}(1+\epsilon_{11})-a_{20}(1+\epsilon_{22})\right]^3\nonumber\\
+b\left[a_{10}(1+\epsilon_{11})-a_{20}(1+\epsilon_{22})\right]+c\left[a_{10}\epsilon_{11}+a_{20}\epsilon_{22}\right],\nonumber
\end{eqnarray}
and:
\begin{eqnarray}\label{eq:eq5}
\frac{1}{a_{10}^2}\frac{\partial^2 U}{\partial \epsilon_{11}^2}=3a\left[a_{10}(1+\epsilon_{11})-a_{20}(1+\epsilon_{22})\right]^2 &-b+c,\nonumber\\
\frac{1}{a_{20}^2}\frac{\partial^2 U}{\partial \epsilon_{22}^2}=3a\left[a_{10}(1+\epsilon_{11})-a_{20}(1+\epsilon_{22})\right]^2 &-b+c,\nonumber\\
\frac{1}{a_{10}a_{20}}\frac{\partial^2 U}{\partial \epsilon_{11}\partial \epsilon_{22}}=-3a[a_{10}(1+\epsilon_{11})-a_{20}(1 &+\epsilon_{22})]^2\nonumber\\+b+c.&
\end{eqnarray}

Additionally, the unit cell area $\mathcal{A}$ can also be parameterized from unitary displacements as:
\begin{equation}\label{eq:eq6}
\mathcal{A}(\epsilon_{11},\epsilon_{22})=a_{10}a_{20}(1+\epsilon_{11}+\epsilon_{22}+\epsilon_{11}\epsilon_{22}),
\end{equation}
with derivatives:
\begin{eqnarray}\label{eq:eq7}
\frac{\partial \mathcal{A}}{\partial \epsilon_{11}}=a_{10}a_{20}(1+\epsilon_{22}),\text{ }
\frac{\partial \mathcal{A}}{\partial \epsilon_{22}}=a_{10}a_{20}(1+\epsilon_{11}),\nonumber\\
\frac{\partial^2 \mathcal{A}}{\partial \epsilon_{11}^2}=\frac{\partial^2 \mathcal{A}}{\partial \epsilon_{22}^2}=0, \text{ and }
\frac{\partial^2 \mathcal{A}}{\partial \epsilon_{11}\partial \epsilon_{22}}=a_{10}a_{20}.
\end{eqnarray}

\begin{figure*}[tb]
\begin{center}
\includegraphics[width=0.96\textwidth]{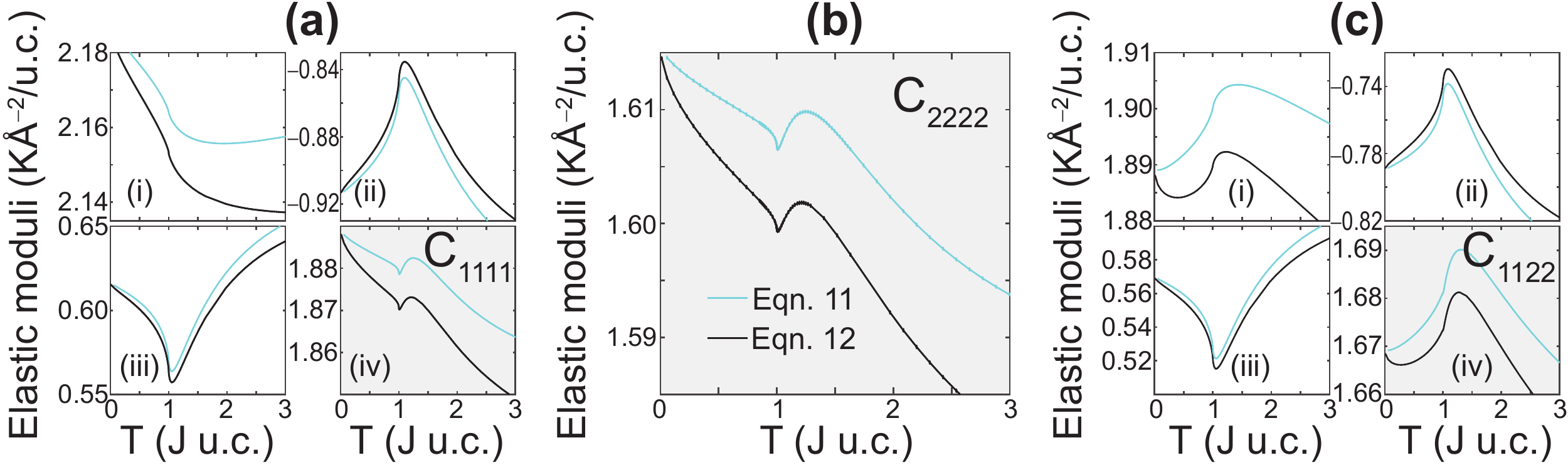}
\caption{(a) The three individual contributions to $C_{1111}$ in Eqns.~\ref{eq:eq11} (cyan) and \ref{eq:eq12} (black) are shown in subplots (i) to (iii) and their sum in subplot (iv). (b) $C_{2222}$. (c) The three individual contributions to $C_{1122}$ in Eqns.~\ref{eq:eq11} (cyan) and \ref{eq:eq12} (black) are seen in subplots (i) to (iii) and their sum is subplot (iv). The sudden downward spikes in $C_{1111}$ and $C_{2222}$ at $T=J$ prove that the analytical model indeed captures the softening of elastic moduli induced by the structural transformation.}\label{fig:f2}
\end{center}
\end{figure*}

Elastic moduli may also be expressible from an energy density $u$, defined for these two-dimensional materials as an energy per unit cell area:
\begin{equation}\label{eq:eq8}
u(\epsilon_{11},\epsilon_{22})\equiv \frac{U(\epsilon_{11},\epsilon_{22})}{\mathcal{A}(\epsilon_{11},\epsilon_{22})},
\end{equation}
which is a small quantity (a discrete ``differential'') within a macroscopic monodomain already (thus not requiring a definition of the type $u=\partial U/\partial \mathcal{A}$).

Elastic moduli are thermal averages. At temperature $T$, Eqn.~$k_BT=U(X,Y)$ has four roots:\cite{arxiv}
$X_{P\pm}(T)=\sqrt{(b\pm\sqrt{4ak_BT})/2a}$, and
$X_{N\pm}(T)=-X_{P\pm}(T)$,
where $N$ ($P$) stands for negative (positive). (In previous expressions, $k_B$ is Boltzmann constant.)

Being a classical construct, the elastic energy profile forbids direct tunneling among the two wells, so one is constrained to ($X_{min}(T)=X_{P-}(T)\le X\le X_{P+}(T)=X_{Max}(T)$) when $k_BT\le J$ at monodomain $A$. Both wells are accessible when $T>J$, and $X_{min}(T)=X_{N+}(T)\le X\le X_{P+}(T)=X_{Max}(T)$. $X_{min}(T)$ takes on two different values, depending on whether $T\le J$ or $T> J$.

This way, the isoenergy contours shown in Fig.~\ref{fig:f1} (which are borne out from the parametrization of the energy landscape, Fig.~\ref{fig:f0}(a), given by Eq.~\ref{eq:U}) are expressed as: $Y_{\pm}(X,T)=\pm\sqrt{(k_BT-aX^4+bX^2-J)/c}$, and ensemble averages within the model for any function $f(X,Y)$ are obtained from\cite{arxiv,Kittel}:
\begin{equation}\label{eq:8}
\langle f(X,Y)\rangle\equiv\frac{\int_{X_{min}(T)}^{X_{Max}(T)} \int_{Y_-(X,T)}^{Y_+(X,T)}e^{-U/k_BT} f(X,Y) dXdY}{\int_{X_{min}(T)}^{X_{Max}(T)} \int_{Y_-(X,T)}^{Y_+(X,T)} e^{-U/k_BT} dXdY}.
\end{equation}
which requires reexpressing $\epsilon_{11}$ and $\epsilon_{22}$ in Eqns.~\ref{eq:eq4} through \ref{eq:eq8} in terms of $X$ and $Y$; something accomplished by inversion of Eqn.~\ref{eq:eq3}.

Two expressions for the elastic moduli were considered:
\begin{eqnarray}\label{eq:eq11}
&C_{ijkl}=\\
&\frac{1}{\mathcal{A}_0}\left\{\left\langle\frac{\partial^2 U}{\partial \epsilon_{ij}\partial \epsilon_{kl}}\right\rangle
-\frac{1}{k_BT}\left[\left\langle \frac{\partial U}{\partial \epsilon_{ij}}\frac{\partial U}{\partial \epsilon_{kl}}\right\rangle
-\left\langle \frac{\partial U}{\partial \epsilon_{ij}}\right\rangle \left\langle\frac{\partial U}{\partial \epsilon_{kl}}\right\rangle\right] \right\}\nonumber,
\end{eqnarray}
where $\mathcal{A}_0=a_{10}a_{20}$, and:
\begin{eqnarray}\label{eq:eq12}
&C_{ijkl}=\left\langle\frac{\partial^2 u}{\partial \epsilon_{ij}\partial \epsilon_{kl}}\right\rangle\\
&-\frac{1}{k_BT}\left[\left\langle \mathcal{A}\frac{\partial u}{\partial \epsilon_{ij}}\frac{\partial u}{\partial \epsilon_{kl}}\right\rangle
-\left\langle\mathcal{A}\right\rangle\left\langle \frac{\partial u}{\partial \epsilon_{ij}}\right\rangle \left\langle\frac{\partial u}{\partial \epsilon_{kl}}\right\rangle\right] \nonumber,
\end{eqnarray}
in which the temperature-induced change of $\mathcal{A}$ is explicitly included in the thermal averages. {Under the assumption that $P$ is near zero (see discussion after Eqn.~\ref{eq:constants}), Eqns.~\eqref{eq:eq11} and \eqref{eq:eq12} are alternative expressions for the second-order derivative of the Helmholtz free energy. The first term to the right of these equations is the average of the second-order derivative of the elastic energy with respect to unitary displacements, while the second and third terms are standard contributions from the system's entropy.}

{We considered the area of the ground state structure in the denominator of Eqn.~\eqref{eq:eq11}, and introduced a variable area into estimations of the average in Eqn.~\ref{eq:eq12}. As seen in Fig.~\ref{fig:f2}, both expressions lead to similar results.}

Subplots (i) to (iii) in Fig.~\ref{fig:f2}(a) are the three contributions to $C_{1111}$. The cyan trends were obtained from Eqn.~\ref{eq:eq11}, while black curves were correspondingly obtained from Eqn.~\ref{eq:eq12}. The explicit display of these three individual terms permits observing their dependence on $T$ and their order of magnitude on the elastic moduli separately. In turn, subplot (iv) shown in grey in Fig.~\ref{fig:f2}(a) displays $C_{1111}$, which is the sum of subplots (i) to (iii).

$C_{2222}$ is shown in Fig.~\ref{fig:f2}(b). The dependency of individual terms on $C_{2222}$ is similar to that observed in Fig.~\ref{fig:f2}(a) and not explicitly shown for that reason. Individual contributions to $C_{1122}$ from Eqns.~\ref{eq:eq11} and \ref{eq:eq12} can be seen in subplots (i) to (iii) of Fig.~\ref{fig:f2}(c), and $C_{1122}$ is shown in Fig.~\ref{fig:f2}(c), subplot (iv).

Results obtained from Eqns.~\ref{eq:eq11} and \ref{eq:eq12} are qualitatively similar. Consideration of the varying area $\mathcal{A}$ onto the thermal averages results in a larger range of change for these elastic moduli. Sudden negative spikes at $T=J$ represent a sudden softening of elastic constants once the two wells on the energy landscape become accessible, as the structural transition onto a square structure takes place. While we will provide a second argument for the barrier height being too small for the two wells to classically confine a chosen domain, the value of the results discussed here and shown in Fig.~\ref{fig:f2} rests on them probably representing the first study of a sudden softening of elastic constants at a structural transformation within the context of two-dimensional materials.

\begin{figure}[tb]
\begin{center}
\includegraphics[width=0.48\textwidth]{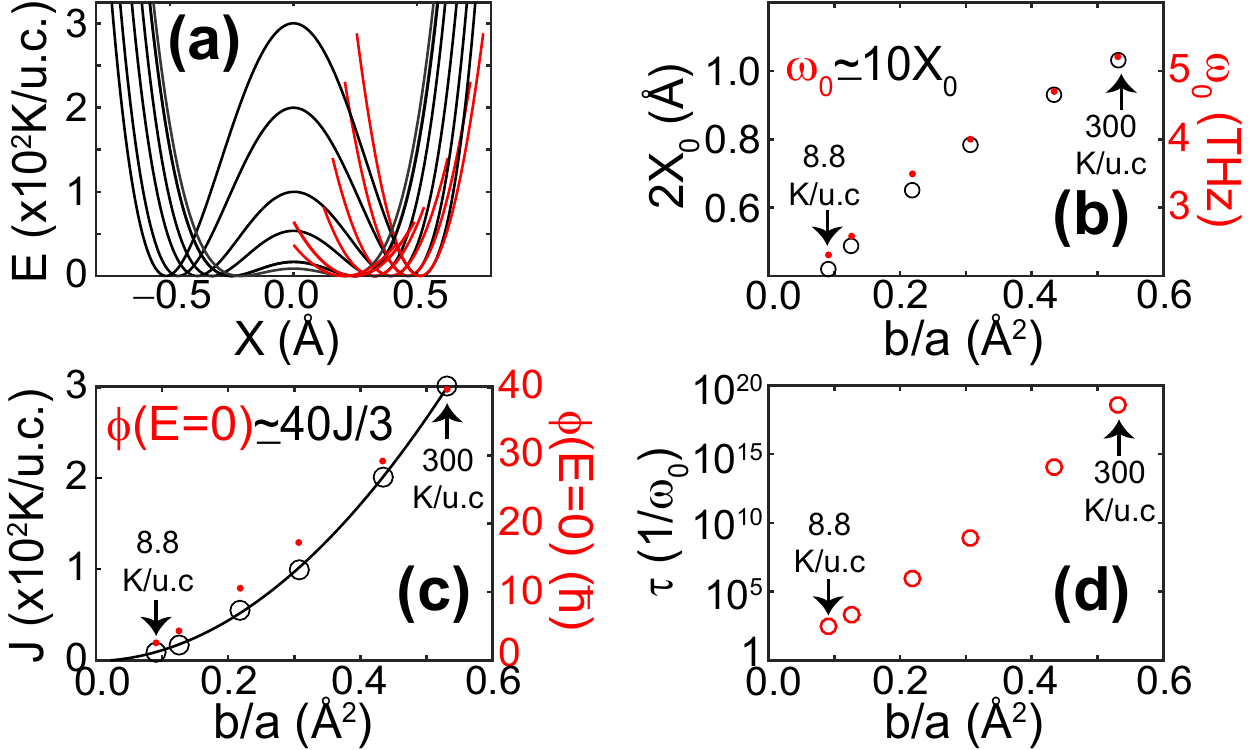}
\caption{(a) The energy profile $U(X,O)$ for six increasing values of the parameter $b$. Parabolic fittings are shown in red. Evolution of (b) wall minima $X_0$ and frequency out of quadratic wells $\omega_0$, (c) elastic energy barrier $J$ and phase $\phi$, and of the escape time $\tau$ as a function of the parameter $b/a$.}\label{fig:f3}
\end{center}
\end{figure}


\section{Escape times as a second argument for paraelastic behavior}\label{sec:3}
We wish to employ the WKB approximation, as discussed in elementary Quantum Mechanics,\cite{griffiths} to estimate the escape times from the double well, considering it as one-dimensional given the steepness of the elastic landscape along the $Y-$direction. The process is not intended to be quantitative, but it will make qualitative sense and will support the hypothesis of a quantum paraelastic phase  given recently.\cite{arxiv}

Considering a particle at the bottom of the well with mass $m=2m_{Sn}+2m_{O}$ ($m_{Sn}$ is the mass of a tin atom and $m_O$ that of an oxygen atom), the process is accomplished in three steps (c.f. pages 336--338 in Ref.~\onlinecite{griffiths}):
\begin{enumerate}
\item To approximate the (order four) double well potential into two square potentials centered at each of the two wells:
\begin{equation}\label{eq:approx}
V(X) \to \begin{cases}
           m\omega_0^2(X+X_0)^2/2, & \mbox{if } X<0 \\
           m\omega_0^2(X-X_0)^2/2, & \mbox{if } X>0,
         \end{cases}
\end{equation}
such that an estimate of the oscillation frequency $\omega_0$, valid near the bottom of the well, can be extracted.
\item To use $U(X,0)$ to estimate a WKB ``phase factor'' at the bottom of the well ($E=0$):
\begin{equation}\label{eq:WKB1}
\phi=\frac{1}{\hbar}\int_{-X_0}^{X_0}|p(X)|dX,
\end{equation}
with $|p(X)|=\sqrt{2mU(X,0)}$, and $U(X,0)$ from Eqn.~\ref{eq:U}.\cite{arxiv,Seixas2016}

\item{} To estimate the escape time $\tau$ from the bottom of one well onto the opposite well via:
\begin{equation}\label{eq:escape}
\tau=\frac{2\pi^2}{\omega_0}\exp[\phi],
\end{equation}
\end{enumerate}
where $\omega_0=2.31$ THz, $\phi=2.81$, and an escape time of only $\tau=1.4\times 10^{-10}$ s, which implies a probability of tunneling among both wells at a rate of 10$^{10}$ per second, indicating that individual wells are not confining for these values of $a$ and $b$.

\begin{figure}[tb]
\begin{center}
\includegraphics[width=0.48\textwidth]{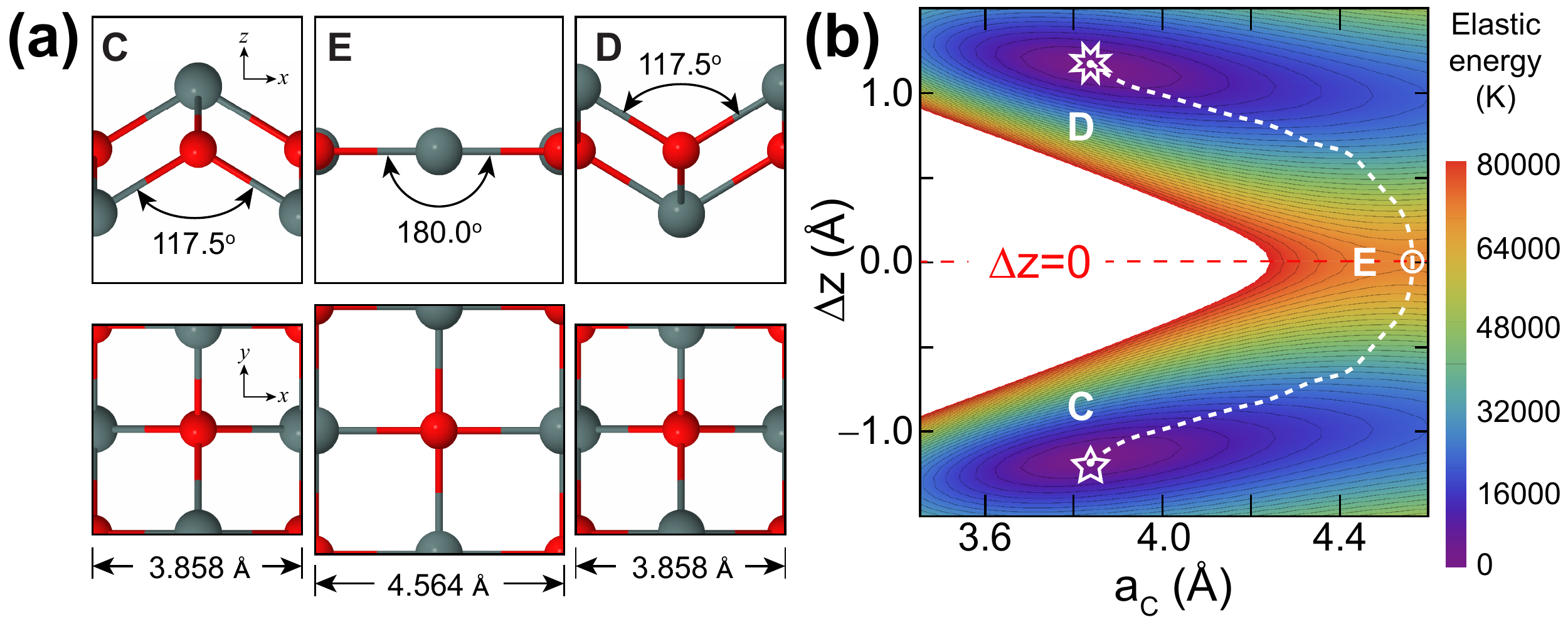}
\caption{(a) A second possible structural transformation, in which the litharge structure turns onto a square lattice. (b) Two dimensional landscape for a SnO monolayer past the rectangular to square transformation. The second energy barrier $J'$ is the energy difference among degenerate points $C$ or $D$ and point $E$, representing a planar square.}\label{fig:f4}
\end{center}
\end{figure}

In Fig.~\ref{fig:f3}, we kept $a=4252$ K\AA$^{-2}$/u.c. and thought of $b$ as a parameter in order to study the magnitude of the escape time as a function of the analytical landscape. We assigned the following values to $b$: 387, 534, 983, 1305, 1844, and 2260  K\AA$^{-2}$/u.c. This way, $b$ increases by 5.8 from the selected lower to the upper limits, raising $J$ from 8.8 K/u.c. to 16.8, 54.0, 100.0, 200.0, and 300.0 K/u.c., respectively, which implies a 34-fold increase of $J$ in between end values for $b$.

The increase in $b$ in these models does not affect the oscillation frequency $\omega_0$ on the square wells shown in red on Fig.~\ref{fig:f3}(a) significantly, whose value changes from 2.3 to 5.2 THz, making for a discrete twofold increase. In Fig.~\ref{fig:f3}(b), one observes a relation among $\omega_0$ and $X_0$, which indicates that the distance among the bottom of the two wells also increases twofold in going from $b=387$ to 2260 K/(\AA$^2$u.c.). The phase factor $\phi$ in turn changes from 2.8 onto 39.7, and Fig.~\ref{fig:f3}(c) one observes an empirical relation $\phi\simeq 40J/3$.

{Fig.~\ref{fig:f3}(d) shows the main result of this section.} Namely, that a sixfold increase on $b$ makes the escape time rise by 15 orders of magnitude, while the barrier $J$ only increases from 8.8 K/u.c. to 300 K/u.c. {At $J=8.8$ K/u.c., the escape time is so short, that it cannot be assumed that a ``particle'' can stay long at an individual well (monodomain), implying once again the quantum paraelastic behavior alluded for in Ref.~\onlinecite{arxiv}.}

\section{No additional two-dimensional transition}\label{sec:4}

The structural transition discussed in previous work\cite{Seixas2016,arxiv} turns a rectangular unit cell with lattice constants $a_1 > a_2$ onto a square with side $a_c$ in which two oxygen atoms lie on a plane, and two tin atoms are at a relative height $\Delta z$ or $-\Delta z$, respectively with $\Delta z>0$. {As the final point to make in the present work, one can envision a second structural transition onto a higher symmetry structure having $\Delta z=0$ shown in Fig.~\ref{fig:f4}(a)}, in which the degenerate states $C$ and $D$ transition onto (an average planar) structure $E$. Such a second transition requires a huge amount of energy nevertheless. {Turning the angle among oxygen and two tin atoms from 117$^{\circ}$ onto 180$^{\circ}$, and the lattice constant from 3.858 \AA{} into 4.564 \AA requires overcoming an energy barrier $J'$ along the dashed path in Fig.~\ref{fig:f4}(b) of the order of 64,000 K/u.c.; such a high magnitude for $J'$ implies that the SnO monolayer melts rather than undergoing such a second two-dimensional structural transformation.}

\section{conclusions}
To conclude, the softening of elastic constants has been discussed within the context of engineering structures and soft matter such as dilute lattices, jammed systems, biopolymer networks and network glasses. Here, it makes its way into the realm of two-dimensional materials, for which exciting additional quantum-mechanically driven interplays are to be expected. We facilitated an incipient procedure to estimate the elastic moduli using an analytical expression for the energy landscape, calculated escape times out of one of the two wells as an additional argument towards a non-negligible quantum tunneling when the energy barrier $J$ is of the order of 10 K per unit cell, and provided arguments against a subsequent two-dimensional structural transition in which the unit cell turns from the slightly buckled litharge structure onto a planar square lattice. {Taken together, these results enhance the toolset to study structural transformation in two-dimensional materials beyond graphene.}

\acknowledgments
A.P.S. is funded by FONDECYT, project No 1171600 (Chile); T.B. by the National Science Foundation (Grant No. DMR-1610126), S.B.L. by the U.S. Department of Energy, Office of Basic Energy Sciences, Early Career Award DE-SC0016139. Part of this work was performed at the Center for Nanoscale Materials at Argonne National Laboratory, a U.S. Department of Energy Office of Science User Facility, and supported by the U.S. Department of Energy, Office of Science, under Contract No. DE-AC02-06CH11357. Conversations with P. Darancet, W. Harter, G. Naumis {and J. W. Villanova} are gratefully acknowledged.


%

\end{document}